\documentclass[preprint]{acmart} 
\usepackage{float}

\setcopyright{none}

\acmYear{2018}
\copyrightyear{2018}

\newcommand{\squeezeup}{\vspace{-6mm}}

\begin{document}
\title[Controlling distilleries in fault-tolerant quantum circuits]{Controlling distilleries in fault-tolerant quantum circuits: problem statement and  analysis towards a solution}

\author{Alexandru Paler}
\orcid{0002-1536-8858}
\affiliation{%
  \institution{Johannes Kepler University, 4040 Linz, Austria}
  \department{Linz Institute of Technology \& Institute Integrated Circuits}
}
\email{alexandru.paler@jku.at}

\renewcommand{\shortauthors}{A. Paler}

\begin{abstract}
The failure susceptibility of the quantum hardware will force quantum computers to execute fault-tolerant quantum circuits. These circuits are based on quantum error correcting codes, and there is increasing evidence that one of the most practical choices is the surface code. Design methodologies of surface code based quantum circuits were focused on the layout of such circuits without emphasizing the reduced availability of hardware and its effect on the execution time. Circuit layout has not been investigated for practical scenarios, and the problem presented herein was neglected until now. For achieving fault-tolerance and implementing surface code based computations, a significant amount of computing resources (hardware and time) are necessary for preparing special quantum states in a procedure called distillation. This work introduces the problem of how distilleries (circuit portions responsible for state distillation) influence the layout of surface code protected quantum circuits, and analyses the trade-offs for reducing the resources necessary for executing the circuits. A first algorithmic solution is presented, implemented and evaluated for addition quantum circuits.
\end{abstract}




\maketitle

\section{Introduction}

The execution of quantum computations is preceded by the preparation of corresponding quantum circuits, which have to be adapted to the availability of the underlying quantum hardware. From a practical perspective, the first generation of quantum computers is envisioned to be constructed as a two-dimensional array of quantum hardware (physical qubits) which will be interacted for the duration of the circuit's execution. The amount of available computational resources is determined by the \emph{area} of the two dimensional physical qubit arrangement (number of physical qubits) and the computation execution time. Therefore, it is usual to refer to the computational resources space-time \textit{volume} (area $\times$ time) \cite{fowler2012bridge,bishop2017quantum}.

There are two issues related the preparation of quantum circuits. First, quantum computations are described in a non-fault-tolerant manner, such that these have to be initially compiled into fault-tolerant circuits \cite{paler2017fault}. Second, fault-tolerance is not only a structural property of the circuits, but also the result of using quantum error correcting codes (QECCs): fault-tolerant circuits are applications of error corrected (logical) quantum gates on error corrected (logical) qubits. The error threshold tolerated by a QECC and the complexity of implementing the QECC in hardware and in software are the criteria determining the QECC's practicality. From this perspective, it is increasingly believed that the surface QECC is the most viable alternative for the first generation of large scale error corrected quantum computers. Accordingly, non-fault-tolerant quantum computations have to be compiled into surface code protected quantum circuits in order to allow their execution.

The second preparation issue is the fact that any QECC implementation necessitates a large number of physical gates and physical qubits. On the one hand, for the surface code, those numbers are realistically smaller than for any other quantum error correcting code possessing a similar threshold. On the other hand, the first quantum computers will have limited hardware, such that quantum circuits compilation has to be efficient for a very restricted environment. Circuit layout possibilities are limited and determined by two factors: 1) physical hardware availability; and 2) fault-tolerant quantum circuit structural properties.

\begin{figure}[h!]
\centering
\includegraphics[width=\columnwidth]{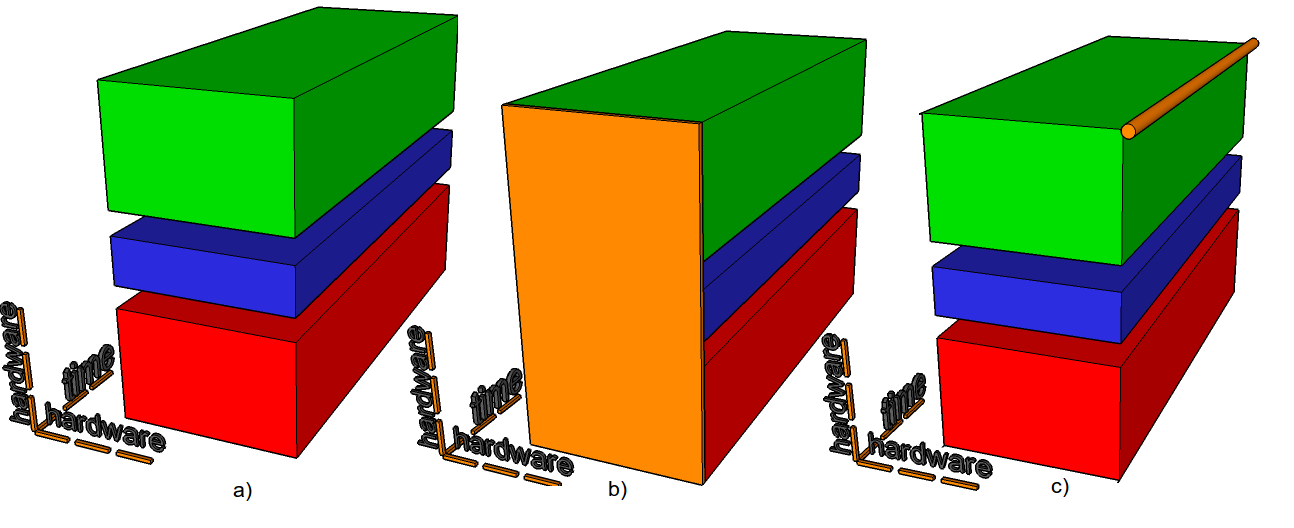}
\squeezeup
\caption{3D Layout: a) bounding box of distillery (green), connections pool (blue) and computation (red); b) necessary hardware resources are abstracted by the area of the orange rectangle; c) execution time is indicated by the length of the orange line. The volume of a layout is area $\times$ time.}
\label{fig:s123}
\end{figure}

This work presents a method for controlling the dimensions of the layout of fault-tolerant quantum circuits (e.g. in Fig.~\ref{fig:s123} the width of the blue cuboid influences the overall bounding box -- the necessary computational resources to execute the circuit associated with the layout). An extensive visually supported introduction to the problem statement is presented in the following. The results are analysed and the algorithmic solution is described. Finally, conclusions and future work are formulated.

A purpose of this work is to emphasize that in the case of fault-tolerant quantum circuits there are classical research problems, which can abstract away quantumness of the circuits. Consequently, this section introduces the problem's statement without needing technical details related to the surface QECC. An exhaustive and clear exposition of surface QECCs can be found in \cite{FMM13}.

\subsection{Background}
\label{sec:back}
\newcommand{\scqc}{\texttt{scqc} }

\begin{figure}[h!]
\centering
\includegraphics[width=0.3\columnwidth]{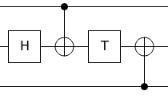}
\squeezeup
\caption{Example: Three qubit quantum circuit. Horizontal lines represent qubits, the two qubit CNOT gate is represented by the vertical line connecting the $\bullet$ to $\oplus$, and the T and H gates are single qubit gates.}
\label{fig:qcirc}
\end{figure}

Quantum circuits consist of quantum wires and a time ordered sequence of quantum gates. Each quantum wire (drawn horizontally in the diagrams) is an abstraction of a qubit (e.g. Fig~\ref{fig:qcirc}). Considering an imaginary time axis running from the left to the right of the quantum circuit diagram, circuit inputs are on the left hand side of the diagram, and outputs on the right hand side. Because quantum computations are reversible \cite{NC00}, circuits have an equal number of inputs and outputs. Quantum gates are applied to a single or multiple qubits.

Surface code protected quantum circuits (in the following called \scqc) are fault-tolerant through the usage of the surface QECC. There exists a canonical translation between a quantum circuit diagram and the corresponding \scqc \cite{paler2017synthesis}, and the result has a three dimensional layout. This layout can be mapped to the physical quantum hardware \cite{paler2014mapping}, which was mentioned having a two-dimensional arrangement. The third dimension of the \scqc layout, along a time axis, is generated by a clear sequence of hardware interactions. This means that the three dimensional layout is sliced along the time axis, and the hardware executes at a given time step the corresponding slice. After mapping, each slice is represented by a sequence of physical qubits and physical gates (belonging to the underlying hardware and not to be confused with the one from the computation/circuit). As a result, it is said that the \scqc operates at a logical layer, where logical elements are constructed through sets of physical elements. The state of the art methods for obtaining \scqc are introduced in \cite{paler2017synthesis, javadi2017optimized}.

\begin{figure*}[h!]
\centering
\includegraphics[width=0.9\textwidth]{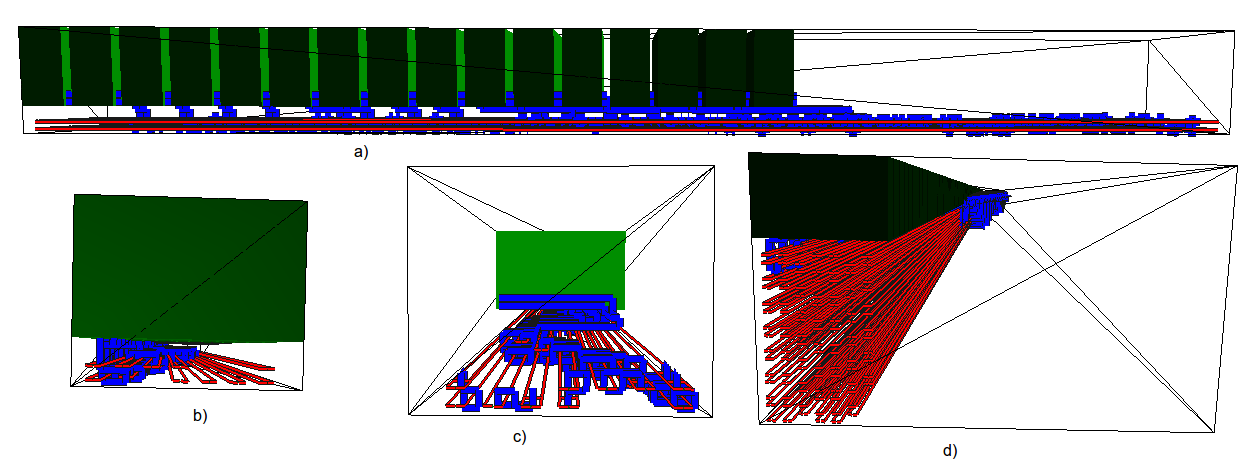}
\squeezeup
\caption{More detailed 3D layouts. The colour coding from Fig.~\ref{fig:s123} was maintained. The bounding box is the black wire-frame. a) A surface code protected quantum circuit. Time runs from left to right. Each green box is a distillation. b) Front view (towards inputs, time runs front to back) Qubits (red) are arranged underneath the distillation boxes; c) Back view (towards outputs, time runs back to front). The circuit's end does not require distilled T states.; d) Too many connections in the pool increase the bounding box.}
\label{fig:3d}
\end{figure*}

Surface code elements are very briefly introduced, and without loss of generality, technicalities of the surface QECC (e.g. planar \cite{FMM13} or defect logical qubits \cite{raussendorf2007topological}) are not discussed. Logical qubits in an \scqc are represented by pairs of lines, and there are two types of lines which can be braided to perform a logical CNOT on the associated logical qubits. Lines are coloured red and blue (e.g. Fig.~\ref{fig:3d}). Lines of the same type can also be braided, but this action leaves the qubit states unchanged. Except braiding there are not many options for performing logical gates (see the following paragraphs), and the CNOT gate is not quantum computing universal. However, universality can still be achieved by using special states (called magic states) \cite{bravyi2005universal}, which are encoded in logical qubits. It has been shown that a fault-tolerant circuit which is universal has a very regular structure: single qubit initialisations (including in magic states), CNOT gates and single qubit measurements \cite{paler2017fault}. 

Thus, one can be prepare an \scqc even if the surface QECC supports only the CNOT natively. However, the difficulty is that logical qubits cannot be initialised with perfect magic states (they are not accurate: deviate from the ideal state necessary for fault-tolerant computation). The advantage of these states is that they can be \emph{distilled} \cite{FMM13}: multiple instances of the same state can be input to a dedicated circuit (\emph{distillation circuit}) that outputs a more accurate (closer to the ideal state) magic state. Executing multiple layers of distillation (multiple distilled states are distilled again in a distillation circuit) polynomially increases the accuracy of the states. The states can be used once it is assumed that their accuracy is sufficient not to break the fault-tolerance of the computation.

There are multiple types of distillable magic states \cite{bravyi2005universal}, but only one type is required for implementing the Clifford+T gate set. The T state (previously it was often referred to as the A state) is used to implement the T gate in the fault-tolerant circuits. The Clifford gate set (of which the CNOT is a member) can be efficiently simulated on a classical computer, and it is the T gate that exponentially increases the simulation complexity. It is possible to implement all the gates from the Clifford set without the use of magic states code by using non-canonical methods \cite{brown2017poking} (which were formulated recently, and not in the initial work \cite{raussendorf2007topological}). Nevertheless, for each T gate a distilled T state is necessary, and each distilled state is obtained by its corresponding distillation circuit. The number of T gates (called \emph{T-count}) in a circuit determines how many distillation sub-circuits have to be included. Therefore, the research efforts of lowering the T-count from arbitrary (e.g. \cite{heyfron2017efficient}) or specific circuits (e.g. \cite{gidney2017halving}) is not relevant only for simulating larger quantum circuits classically, but also for decreasing the number of additional resources (number of distillation sub-circuits) necessary for the execution on a quantum computer.

\subsection{Circuit Layout}

A \scqc layout is three-dimensional (Fig.~\ref{fig:s123}) and its bounding box should be smaller or at the most the same size of the bounding box of the available resources, otherwise the \scqc cannot be executed. A first aspect is that the bounding box side next to the inputs should fit into the two-dimensional physical qubits area (illustrated orange in Fig.~\ref{fig:s123}b)). Execution time is not limited, and the longest side of the \scqc bounding box is not limited either (Fig.~\ref{fig:s123}c)). A second aspect is that one expects to execute a computation as fast as possible. The first aspect cannot be controlled for arbitrary quantum computations (some computations are simply too large to be executed), but it is possible to control the second one.

This work introduces the problem of how distillation sub-circuits and their arrangement in the \scqc layout influences the computational resource volume (Fig.~\ref{fig:3d}a). We restrict the problem to a realistic scenario: the first generation of quantum computers will have a limited amount of available hardware, while execution time will not be limited. Because of this scenario, the \scqc layout used herein is a simplification of the one presented in \cite{paler2017synthesis}.

The next sections present the layout in top-down manner, and discuss the decisions which led to this design. The bounding box of a \scqc layout (e.g. Fig.~\ref{fig:3d}) has three partitions: 1) distillery; 2) connections pool; 3) computation (Fig.~\ref{fig:s123}).

\begin{figure*}[t]
\centering
\includegraphics[width=0.85\textwidth]{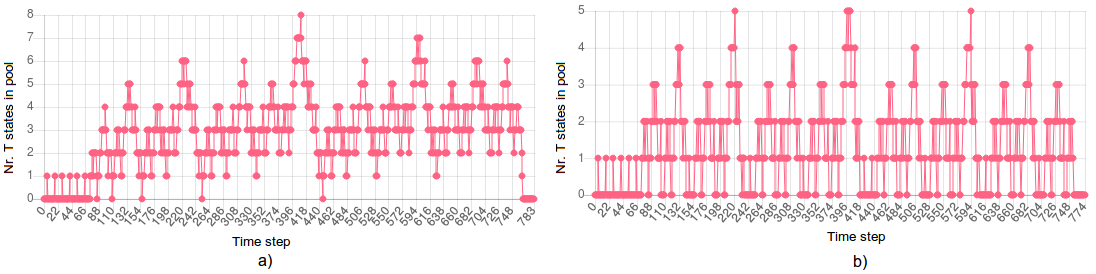}
\squeezeup
\caption{Example: Number of available distilled T states in a pool (vertical axis). Both figures are plotted for the same gate list.  The different plots were obtained without controlling the distillery (start \& stop), but by arranging the gates in different manners. The circuit from a) has a depth (horizontal axis, longest path in the circuit connecting an input to an output) of 783, while in b) the depth is 774. The plot a) was generated by allowing multiple T gates in the computation to be executed in parallel, and b) does not allow this (single T gate per time step). The current qubit arrangement in the computation partition does not support parallel T gates.}
\label{fig:avail}
\end{figure*}

\subsection{Distillery}

The distillery is the region in which the distillation sub-circuits are placed. In Fig.~\ref{fig:3d} the distillations are abstracted through \emph{distillation boxes} (green coloured boxes). All green boxes have the same dimensions, because only T states need to be distilled. The currently most compact single level distillation circuit is presented in \cite{fowler2012bridge}, and using that circuit as a basis it is possible to determine the dimensions of multi-level distillations, too.

The number of distillations performed by the distillery is a function of the circuit's T-count. This work assumes that each T gate will require a single distillation (three distillation layers are executed within each distillation box). Each distillation outputs a distilled magic state to be used during the computation. It can be noticed that, in the presence of distillations, the minimum time necessary to execute a circuit is $t_{min}=Tc \times dist_t$, where $Tc$ is the T-count and $dist_t$ the time necessary for each distillation.

\subsection{Connections pool}
\label{sec:pool}

The second partition is where the distilled T states are stored until they are required in the computation (blue lines in Fig.~\ref{fig:3d}). They are called connections, because they function like connections between the distillations and the places in the \scqc where they are necessary. The connections pool is placed between the distillery and the computation partition (Fig.~\ref{fig:s123}a)), and works similar to a wiring duct: it collects the T states and delivers them to the computation. However, the connections are managed like pool resources, because all T states are identical, and any available distilled state could be used for any circuit T gate.

The pool has a planar arrangement/layout (see Fig.~\ref{fig:3d}d)), meaning that connections are placed in a single row whose length is along the time axis, and the row width occupies hardware resources. The pool width is determined by the number of stored connections. The minimum width is zero (no available distilled T states), and the maximum is theoretically unlimited, but practically limited to the width of the distillation boxes (this will be later illustrated).

\subsection{Computation}
\label{sec:comp}

The third part of the layout represents the \scqc equivalent of the quantum circuit diagram that was translated to surface QECC elements. This region includes the set of qubits (pairs of red lines in Fig.~\ref{fig:3d}) with their corresponding inputs and outputs, the single qubit Clifford gates and the CNOTs. Each time a T gate has to be executed, a T state is \emph{consumed} from the connection pool by extending the connection (representing in fact an ancilla qubit of the T state) to the place in the computation where it is needed.

If one would decide, in the very unlikely situation, to execute on a quantum computer only a Clifford circuit (T-count is zero), then only the computation partition would be included in the \scqc.

For resource efficiency reasons, the computation region is organised into rows and columns: the line pair of each qubit is placed at predetermined matrix coordinates: a known row and column. The number of columns is calculated from the width of the distillation boxes, and correspondingly many rows are used to hold all the qubits. For example, considering that there are 77 qubits and 7 columns available, the qubit placement will require 11 rows. Thus, the bounding box of the computation region is a cuboid having the same width as the distillation boxes, and the depth at least as long as the distillery: it can happen that towards the end of the computation no states need to be distilled (no T gates are left in the circuit), and the remaining part of the computation consists entirely of Clifford gates (e.g. Fig.~\ref{fig:3d}c).

\begin{figure*}[t]
\centering
\includegraphics[width=0.9\textwidth]{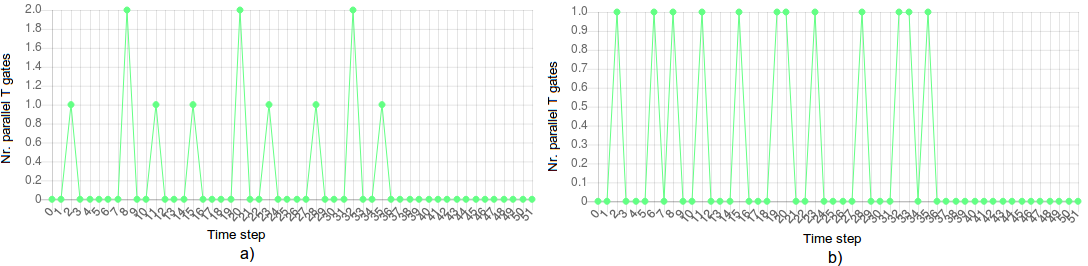}
\squeezeup
\caption{Example: Distribution of T gates in a circuit. The discussion of Fig.~\ref{fig:avail} mentioned parallel T gates. In a) it can be seen that for specific quantum adders \cite{gidney2017halving} the maximum number of parallel T gates is 2 (vertical axis). However, it is possible to rearrange the gates, such that a single T gate is executed at a time step, without affecting the circuit depth (horizontal axis, maximum value is 51).}
\label{fig:distr}
\end{figure*}

\subsection{Problem Statement}

As previously mentioned, the bounding box of the \scqc has three dimensions with one being time (depth), that advances with each slice of the executed \scqc (see Section~\ref{sec:back}). The distillery executes a sequence of distillations with a maximum frequency dictated by the depth (duration) of the distillation boxes. Without controlling the distillery (when to start and stop), connections are added to the pool at a constant frequency.

\begin{figure}
\includegraphics[width=0.8\columnwidth]{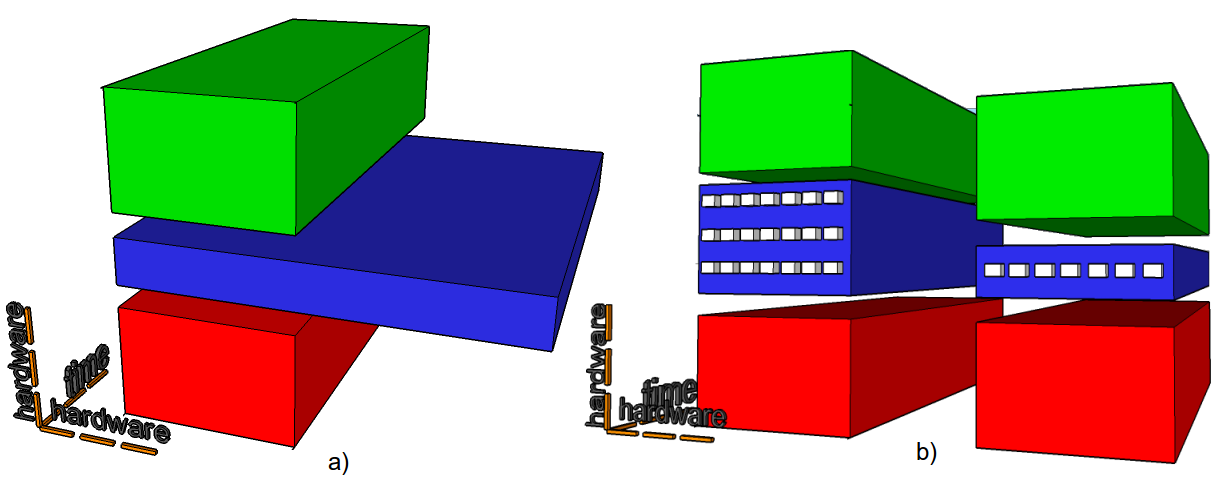}
\squeezeup
\caption{Too many connections in the pool (blue) increase the layout's bounding box. a:) the arrangement is a row; b) the arrangement is a matrix (white bars represent the connections).}
\label{fig:s4}
\end{figure}

The frequency with which connections are consumed is ruled by the occurrence of T gates in the computation (e.g. Fig.~\ref{fig:distr}). On the one hand, if the computation does not require any T state for longer periods of time, the pool's size increases. On the other hand, if many T states are consumed during a short time period, the pool may be emptied and the computation will have to wait (be \emph{delayed}) until new states are distilled and added. A visual example is Fig.~\ref{fig:s4}.

The problem can be stated straightforward: \emph{Control the bounding box of an \scqc by controlling the distillation frequency}. The same problem is equivalent to the following \emph{Devise a distillery start \& stop strategy depending on the number of stored connections in the pool}. Thus, hereinafter the control mechanism will be steered by the maximum capacity of the connection pool.

\subsection{Different Connection Pool Layout}

The connection pool is arranged in a row (see Sec.~\ref{sec:pool}), and this is different from the computation partition which has a matrix arrangement (see Sec.~\ref{sec:comp}). It can be argued, that it should be possible to have the pool as a matrix too, in order to increase the maximum number of available connections. The effect would be that computations would not need to be delayed, and the distillery would not be stopped and started often.

The matrix arrangement is a possibility, but it should be noted that this would require a more complex analysis of the necessary pool dimensions in conjunction with the circuit structure (see Fig.~\ref{fig:distr}). Otherwise, for long periods of time, the connections would remain unused, and synthetically increase the overall \scqc bounding box without a real benefit to the computation. We consider such an analysis a refinement of the problem statement, that will be tackled with more advanced tools (see Sec.~\ref{sec:concl}).

Furthermore, given that the hardware resources are limited, it is more efficient to include the absolute minimum number of distilled T state in the connection pool. The row arrangement is considered the most viable compromise for the first \scqc generation.

\section{Results}

\begin{figure}
\centering
\includegraphics[width=0.9\columnwidth]{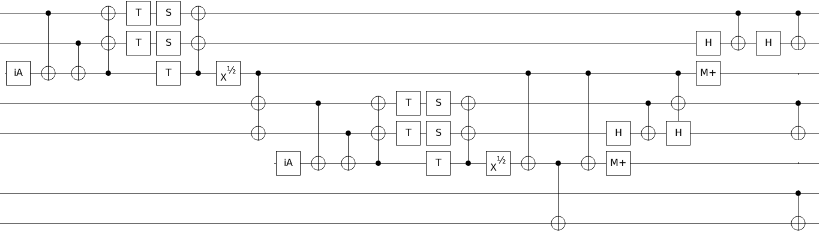}
\squeezeup
\caption{A 3 qubit adder. The \emph{iA} rectangles represent initialisations of qubits into a distilled T state. Those initialisations are included into the T-count. This work considers that T gates cannot be executed in parallel.}
\label{fig:adder}
\end{figure}

The results obtained for preparing the \scqc versions of the quantum adders introduced in \cite{gidney2017halving} are presented in this section (see Table~\ref{tbl:res}). These particular circuits were chosen, because they require only $4n + \mathcal{O}(1)$ T gates for an $n$ qubit addition. Additionally, the adders (or adder-like circuits) are very often used as sub routines in larger quantum algorithms. Fig.~\ref{fig:adder} is the Clifford+T representation of a three qubit adder (the circuit includes two ancilla qubits such that it operates on $3 + 3 + 2 = 8$ qubits).

For the purpose of clarity, the metric used to express the dimensions of the \scqc bounding boxes was not mentioned until now. The \emph{equivalent volume} was introduced in \cite{fowler2012bridge}, and the results in Table~\ref{tbl:res} represent the number of \emph{plumbing pieces} inside the volume. Each plumbing piece can be visualised as a three dimensional cubic unit: a convenient code distance independent measure of space-time volume.

\begin{figure*}[t]
\centering
\includegraphics[width=0.9\textwidth]{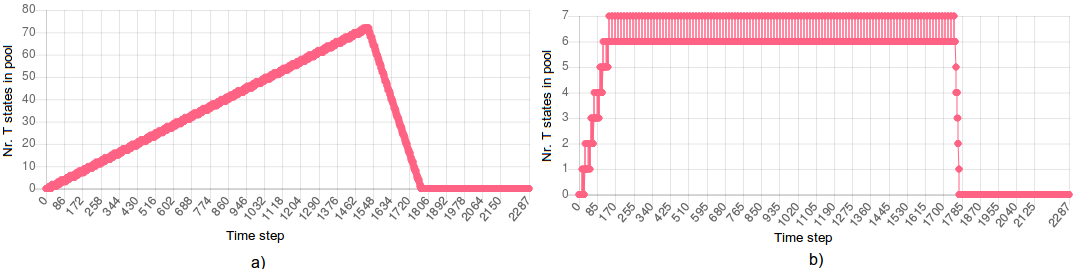}
\squeezeup
\caption{Number of available T states in the pool of the 128 qubit adder. a) Without distillery control; b) With distillery control.}
\label{fig:addpool}
\end{figure*}

The adder's structure is such that all the T gates appear in the first two thirds of the circuit (e.g. Fig.~\ref{fig:addpool}a), and towards circuit outputs only Clifford gates are executed. Because the T gates are not sufficiently close to one another (time distance between two gates is larger than the duration of a single distillation), the pool size increases constantly until all T states have been distilled.

\begin{table}
\begin{tabular}{r | r | r | r | r | r | r | r | r | r}
N & T-Count &  \multicolumn{3}{|c|}{No Control} & \multicolumn{3}{|c|}{Control} & Improv.\\
& & d & w & h & d & w & h & \\
\hline
64 & 252 & 2272 & 73 & 40 & 2272 & 16 & 40 & 4.56 \\
128 & 508 & 4576 & 145 & 67 & 4576 & 16 & 67 & 9.06\\
256 & 1020 & 9184 & 293 & 122 & 9184 & 16 & 122 & 18.31\\
512 & 2044 & 18400 & 585 & 232 & 18400 & 16 & 232 & 36.56
\end{tabular}
\caption{Resource estimation (depth, width, height of the bounding box) for $N$-qubit (first column) quantum adders with and without the control mechanism for the distillery. The very high volumes for the uncontrolled distillery result, because the connection is much wider than the rest of the \scqc layout (see Fig.~\ref{fig:addpool} for comparison).}
\label{tbl:res}
\squeezeup
\squeezeup
\end{table}

Table~\ref{tbl:res} enumerates the resource estimations for quantum adders for 64, 128, 256 and 512 qubits. For example, the 128 qubit adder has a T-count of 508 (see Table~\ref{tbl:res}), and the maximum number of available connections in the pool is 72 at time step 1527. The pool is empty at time step 1772. Between time steps 1527 and 1772, 72 T gates were executed. Thus, until time step 1500, $508 - 72$ T gates were performed, meaning that the a T state was consumed on average every approx. $1527/(508-72)=3.50$ time steps, and after the pool included the maximum connections, the average time was approx. $3.40$. It can be concluded that T gates are required on average at a speed slower than the one of the distillery (a T state every 3 time steps).

After limiting the maximum number of connections in the pool to $7$ (see Sec.~\ref{sec:methods}), none of the adders was delayed: the circuit depth was left unchanged (cf. columns No Control/d and Control/d). With or without the distillery control, the resulting \scqc bounding boxes had the same depth. Nevertheless, the overall hardware necessary to execute each \scqc is improved by significant factors (cf. column Improv.). Figure~\ref{fig:3d}d sketches how the bounding box of the entire computation is increased by the connection pool.

The results, the figures and the plots were automatically generated using tools developed by the author \footnote{https://github.com/alexandrupaler/quantjs} \footnote{https://github.com/alexandrupaler/tqec}.

\section{Methods}
\label{sec:methods}

The high-level, extended introduction and problem statement indicated that the connection pool is the key to the solution. The distillery control logic was implemented in a software tool, which is based on a model of the general circuit layout. The tool simulates the state of the pool and the functionality of the distillery.

The distillery and the pool are modeled as state machines. The pool states are \texttt{full}, \texttt{empty}, \texttt{accepting}. The distillery has the following states:\texttt{working}, \texttt{distilled}, \texttt{stopped}, \texttt{start} and \texttt{stop}. A distillation cannot be interrupted by the pool, such that each time a state is \texttt{distilled} the pool is queried if it accepts one more state. If the pool will be \texttt{full} after the current distillation, the distillery goes first into \texttt{stop} and then \texttt{stopped}. Each time a pool connection is used, the pool is \texttt{accepting} and sets the distillery to \texttt{start} if it is not \texttt{working}. The \texttt{empty} state signals to the computation that it needs to delay its gate applications.

The tool takes a Clifford+T gate list as input, and simulates the \scqc preparation step performed by the quantum computer before the \scqc is executed. Therefore, the output of the tool are the estimated resources for the obtained \scqc and another annotated gate list that includes whenever \emph{distillOn} and \emph{distillOff} were performed. It is assumed that the computer generates layouts similar to the ones in this work.

The methodology uses a stack model of the connection pool, and the stack will have a limited capacity. Ideally the pool will have the same width as the distillery. The distillery's width is determined by a distillation box width, where each box has a depth of 6 plumbing pieces (resulting after being compressed using the method from \cite{fowler2012bridge}), width of 16 pieces (distillation uses 15 ancilla) and height of 10 pieces (for three distillation layers). Thus, the distillery is 16 plumbing pieces wide \cite{fowler2012bridge}.

Due to way how gates are implemented in the surface code, it is known that in canonical layouts (e.g.\cite{paler2017synthesis,FMM13}) each CNOT and single qubit Clifford gate occupy a depth (time) of two plumbing pieces. Therefore, there exists a relation between layout depth $l_d$ and non-\scqc circuit depth $c_d$: $l_d = 2c_d$. For example, a circuit with a depth of 4 CNOTs will result in an 8 pieces deep layout. Therefore, a distillation's depth is 6 plumbing pieces and it requires $6/2=3$ time steps.

Each \scqc qubit is defined by line pairs (red or blue), and a line occupies an area of one plumbing piece. Thus, a maximum of 8 qubits can be arranged beneath the distillation boxes. However, the connections from the pool need additional space to be routed to the computation partition. Therefore, we modeled the pool with a maximum capacity of 7 connections and the computation partition with 7 qubits per row. 

\section{Conclusion}
\label{sec:concl}

This work introduced the problem of laying out surface code protected quantum circuits. The difficulty is to use as few hardware resources as possible, without overly delaying the \scqc execution. The concept of distillery was presented in conjunction with the pool where distilled states are stored until needed during the computation. The maximum size of the pool impacts the amount of necessary hardware resources.

A straightforward method based on treating the pool as a stack with limited capacity was presented. Whenever, the stack is full, the distillery is stopped, and it is started whenever the pool signals that it is ready to accept new distilled states.

Future work will focus on more advanced techniques such a network calculus for modeling this problem. Additionally, it will be investigated how computationally equivalent but structurally different Clifford+T circuits can be used to generate to achieve an average incidence of T gates as close as possible to $3$.

\begin{acks}
The author thanks Dr. Austin Fowler for suggesting this problem. The author was supported by the project CHARON funded by the Linz Institute of Technology.
\end{acks}

\bibliographystyle{ACM-Reference-Format}
\bibliography{biblio}

\end{document}